# Atomic scale spectral control of thermal transport in phononic crystal superlattices

D. Meyer[1], V. Roddatis[2,3], J.P. Bange[1], S. Lopatin[4], M. Keunecke[1], D. Metternich[1], U. Roß[3], I.V. Maznichenko[5], S. Ostanin[5], I. Mertig[5], V. Radisch[3], R. Egoavil[6], I. Lazić[6], V. Moshnyaga[1*], and H. Ulrichs[1**]

[1] *Erstes Physikalisches Institut, Georg-August-Universität Göttingen, Friedrich-Hund-Platz 1, 37077 Göttingen, Germany*

[2] *GFZ German Research Centre for Geosciences, Helmholtz Centre Potsdam, Telegrafenberg, 14473 Potsdam, Germany*

[3] *Institut für Materialphysik, Georg-August-Universität Göttingen, Friedrich-Hund-Platz 1, 37077 Göttingen, Germany*

[4] *Core Lab King Abdullah University of Science and Technology, Thuwal 23955, Saudi Arabia*

[5] *Institut für Physik, Martin-Luther-Universität Halle-Wittenberg, D-06120 Halle, Germany*

[6] *Thermo Fisher Scientific (formerly FEI), Achtseweg Noord 5, 5600KA, Eindhoven, The Netherlands*

*e-mail of corresponding author 1: vmosnea@gwdg.de

**e-mail of corresponding author 2: hulrich@gwdg.de

**Keywords**: phononic crystal, perovskite superlattices, thermal transport, HAADF-STEM, iDPC-STEM, octahedral distortions, ab initio DFT calculations, lattice-dynamical model

We present experimental and theoretical investigations of phonon thermal transport in $(LaMnO_3)_m/(SrMnO_3)_n$ superlattices (LMO/SMO SLs) with the thickness of individual layers $m, n = 3 - 10$ u.c. and the thickness ratio $m/n = 1, 2$. Optical transient thermal reflectivity measurements reveal a pronounced difference in the thermal conductivity between SLs with $m/n = 1$, and SLs with

$m/n = 2$. State-of-the art electron microscopy techniques and ab-initio density functional calculations enables us to assign the origin of this difference to the absence ($m/n = 1$) or presence ($m/n = 2$) of spatially periodic, static oxygen octahedral rotation (OOR) inside the LMO layers. The experimental data analysis shows that the effective thermal conductance of the LMO/SMO interfaces strongly changes from 0.3 GW/m²K for $m/n = 2$ SLs with OOR to a surprisingly large value of 1.8 GW/m²K for $m/n = 1$ SLs without OOR. An instructive lattice dynamical model rationalizes our experimental findings as a result of coherent phonon transmission for $m/n = 1$ versus coherent phonon blocking in SLs with $m/n = 2$. We briefly discuss the possibilities to exploit these results for atomic-scale engineering of a crystalline phonon insulator. The thermal resistivity of this proposal for a thermal metamaterial surpasses the amorphous limit, although phonons still propagate coherently.

## I. Introduction

Being able to control thermal transport properties is of fundamental importance for the development of sustainable and energy efficient technologies. A general reason is that, from a thermodynamic point of view, all periodic energy conversion processes are limited by the success of the thermal management. Thus, material research addressing thermal properties [1 – 13] are key for future progress.

Artificial design of metamaterials, like photonic crystals [14], is a popular approach to produce materials with advanced functionality. Aesthetic examples of photonic metamaterials can be found in nature, where, for instance, the wings of butterflies display beautiful colour effects due to periodic grating structures on their surface, giving rise to complex light reflection phenomena [15]. Taking a Darwinian perspective on their functionality, these colour effects increase the butterfly's reproduction rate, for instance by implementing camouflage. From a technical point-of-view, such photonic structures are implemented in so-called photonic crystal glass fibres [16], where a spatial modulation of the refraction index for light is created on length scales comparable to the wave length of the desired photons. In these examples, the typical wave length of visible photons $\lambda \approx 500$ nm is much larger than the lattice constant ($a \approx 0.5$ nm) of the underlying material. Here, the prefix meta is well justified, because the functionality indeed emerges on scales strongly deviating from the microscopic origin of the modulated material property.

In this article, we present our progress in the field of phononic metamaterials [1, 4, 6, 7, 9, 17 – 24]. When targeting sound waves or elastic dynamics with $\lambda \gg a$, a similar separation of scales as in photonic metamaterials can be found. For example, macroscopic periodically structured materials (phononic crystals) can be designed to block, reflect or redirect a certain range of acoustic frequencies [17, 19]. This makes these materials interesting for sound or vibrational insulation [24], and even for seismic insulation [22]. Downscaling spatial feature sizes to a few nanometers allows to address ultrasonic frequencies up to hundreds of gigahertz [18, 23]. However, to address the most ubiquitous material property related to phonons, which is the thermal conductivity, one has to acknowledge that at technologically relevant non-cryogenic temperatures ($T = 300$ K, $k_\text{B}T = 26$ meV, $\omega_{th}/2\pi = \frac{k_\text{B}T}{h} = 6$ THz) the corresponding phonon wave length $\lambda \approx a$. Such a material cannot be called a metamaterial in a strict sense anymore. The challenge in realizing a phononic crystal which addresses thermal transport at room temperature is that it needs to be designed on the atomic scale [1, 2, 3, 6, 8]. In addition, in contrast to the design of conventional metamaterials, not only a specific frequency needs to be addressed, but instead the broad range from $\omega = 0$ up to $\omega_{th}$. Here, we consider (LMO)$_m$(SMO)$_n$ SLs as a thermal phononic crystal. In the last years, this SL material system has attracted scientific curiosity due the observations of interfacial ferromagnetism [25-34] (FM) with T$_C$~180 K. This FM phase was assigned to LMO(top)/SMO(bottom) interfaces by means of polarized neutron scattering [28]. Very recently, we have observed a high temperature FM phase with T$_C$~360 K at the SMO/LMO interfaces [34]. By means of in situ optical ellipsometry, we have correlated its appearance with interfacial charge transfer from the LMO to the SMO layers.

The remaining article is organized as follows. In section II, we briefly introduce the samples and the optical measurement technique. In section III, we present optical reflectivity measurements, from which we determine the thermal transport properties. Further, the origin of a large difference in the volume and interfacial thermal transport properties between SLs with $m/n = 1$ and $m/n = 2$ is clarified by means of transmission electron microscopy observations (section IV) and density functional theory (DFT) calculations (section V). We show that by stacking these correlated electronic materials, the cubic structure in LMO, not existing in bulk LMO crystals at room temperature, can be energetically stabilized in an SL with LMO thickness 2 < m < 10 u.c.. In section VI, a short overview over known spectral properties of LMO and SMO is presented. All findings are combined in a lattice dynamical model (section VII and appendix section X), which provides an instructive explanation for the change in thermal transport properties. In the discussion section VIII, we detail out that, despite a significant impact of phonon scattering at the interfaces in the actual SLs, our model strongly indicates that in all studied SLs the phonon mean free path exceeds the SL period ($\xi > \Lambda$), and thus the transport is coherent. We close with a proposal for a phononic crystal derived from our SLs, whose thermal conductivity almost vanishes.

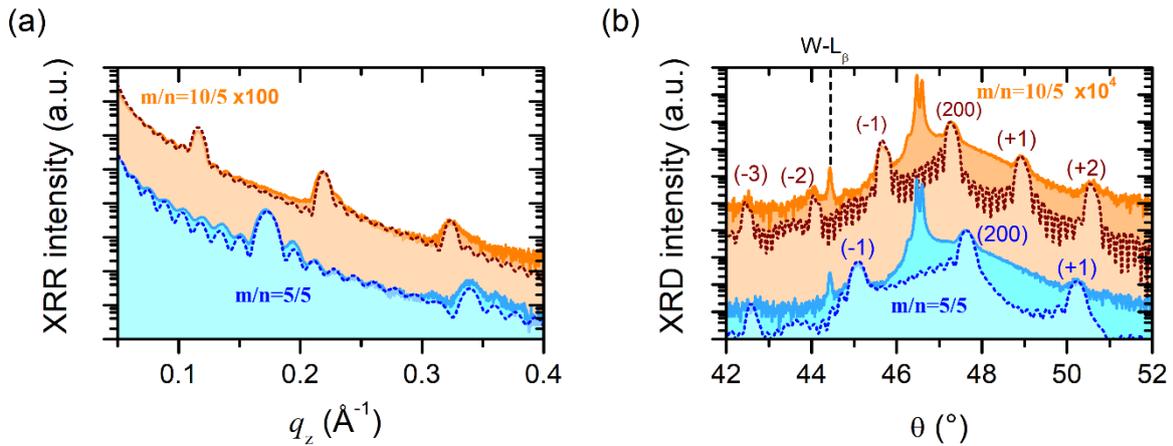

**FIG. 1.** Sample characterization by means of X-ray reflectivity and diffraction. (a) and (b) show XRR and XRD spectra (solid orange and light blue) for two SL samples as indicated. These data were obtained before Cu coverage. Dashed lines are simulations (for XRR simulation see Ref. [35], for XRD simulation see Ref. [36]).

## II. Sample and experimental details

Growth of (LMO)$_m$(SMO)$_n$ SLs with n=3-10 and m/n=1 and 2 on (100)-oriented SrTiO$_3$(STO) substrates has been carried out by a metalorganic aerosol deposition (MAD) technique [34]. Within this method, in situ optical ellipsometry enables atomic layer-resolved control of the thin film growth. X-ray diffraction (XRD) and X-Ray reflectometry (XRR) patterns in Fig. 1 (a) and (b) demonstrate high structural quality of layers and interfaces as well as of the overall SL architecture. The XRR clearly shows Kiessig fringes, associated with the SL period $\Lambda$, and the XRD spectra show satellite peaks corresponding to the SL structure. Fitting these data with the XRR simulation package ReMagX [35] reveals the small root mean-square roughness (*RMS*) of LMO-SMO interfaces between $RMS = 0.2(1)$ nm and $RMS = 0.4(1)$ nm. Further quantitative results are summarized in Table SM-I in the Supplementary Information (SI) [37].

To assess thermal transport properties of the SLs, we performed all-optical thermal transient reflectivity (TTR) experiments [38]. To prepare the SL samples for this method, we deposited a 50 nm thick copper film by electron-beam evaporation. In our implementation of TTR, a nanosecond pulse laser (green, 514 nm) locally heats up the sample, and the resulting temporal evolution of the surface temperature is probed by a cw laser (red, 640 nm). Further details on TTR and the actual setup can be found in the SI [37].

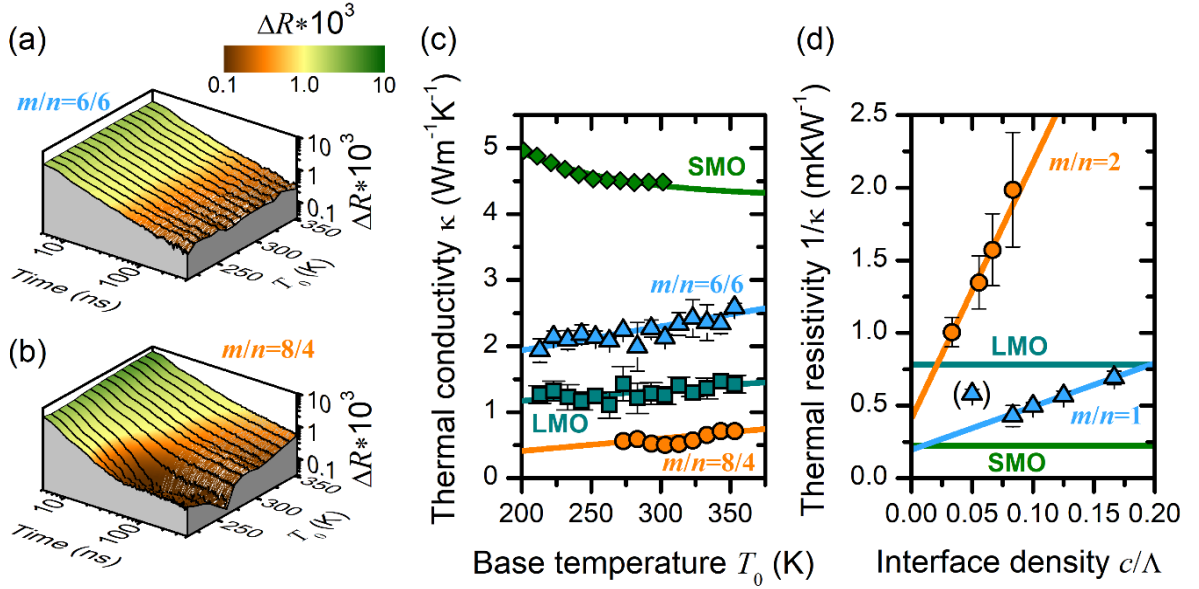

**FIG. 2.** Thermal characterisation of the SLs by means of TTR. (a) and (b) show representative temperature dependent thermal transients for two SLs with m/n as indicated. (c) Thermal conductivity according to a three-layer model [40] for typical SL samples as indicated, as well as for an LMO film. Data for SMO was taken from Reference [40]. Solid lines in the back of the data as a guide to the eye. (d) Thermal resistivity as a function of the interface density for SLs with m/n=1 (blue symbols), and m/n=2 (orange symbols) at $T_0 = 293$ K. The linear trend of these data is fits well to a serial resistance model Eq. (1). Horizontal lines mark the thermal resistivity of LMO (own data) and SMO (from [40]).

### III. TTR results

Figure 2 summarizes the thermal characterization of our SL samples by means of TTR, and the analysis of the data. Representative TTR measurement data for base temperatures from $T_0$=213 - 353 K are shown in Fig. 2(a) (m/n=6/6, SL period $\Lambda = 12c$) and (b) (m/n=8/4, SL period $\Lambda = 12c$); here $c$ is the pseudo-cubic perovskite c-lattice parameter, corresponding to the thickness of 1 u.c.. Generally, one sees a monotonously decaying reflectivity signal. Note that, the slopes of these transients reflect the thermal properties (thermal conductivity $\kappa$, interface conductances $h$, specific heat $c_p$) of the SL, sandwiched in between the Cu capping layer and the STO substrate. A quick look at the data for the SL with m/n=6/6 (see Fig. 2 (a)) reveals that, these parameters do not change significantly as a function of the base temperature $T_0$. For the second sample, which is the SL with m/n=8/4, a weak temperature dependence can be found above 280 K (see Fig. 2(b)). Below 280 K, the dynamics accelerates. This behaviour is likely caused by a significant peak in the specific heat due to a magnetic phase transition. Details about the magnetic properties will be presented elsewhere. Outside of the phase transition, a three-layer (Cu|SL|substrate) thermal model can be applied for the analysis. [39] Further details on this model can be found in the SI to this article [37].

As shown in Fig. 2(c), the analysis quantitatively confirms the observation of a weak temperature dependence of the thermal conductivity. For all our manganite samples, only a weak increase in $\kappa$ can be stated. For comparison, Fig. 2(c) also includes data for SMO taken from [40]. One can see that, our SLs as well as our LMO film have a significantly smaller $\kappa$ than SMO. Moreover, the SL with m/n=8/4 reaches only about a quarter of the thermal conductivity of the SL with m/n=6, which has the same interface density. This strong difference in the thermal conduction is further emphasized in Fig. 2(d), where the thermal resistivity $1/\kappa$ is plotted as a function of the interface density $c/\Lambda$. One sees that samples with m/n=1 and samples with m/n=2 group along two distinct lines, with significantly different slopes. Only the SL with m/n=10/10 does not fit to either trend. To quantify this behaviour, recall that, in a multilayer film, the thermal resistivity can be rationalized within a serial resistance model [1]. In such a picture, the different layers give rise to a volume contribution, and each interface contributes with its interface resistance. In total, we find for our SLs

$$\frac{1}{\kappa} = \frac{1}{\kappa_v} + \frac{2}{h\Lambda}, \qquad (1)$$

where we call $\kappa_v$ the volume thermal conductivity, and $h$ is the interfacial conductance. Fitting Eq. (1) to the data in Fig. 2(d) allows to estimate $\kappa_v$ from the y-intercept, and $h$ from the inverse slope. We find that the SLs with m/n=2 have a volume conductivity of $\kappa_v = 2.4(4)$ Wm$^{-1}$K$^{-1}$, and an interfacial conductance $h = 0.30(3)$ GWm$^{-2}$K$^{-1}$. In contrast, the SLs with m/n=1 possess a volume conductivity of $\kappa_v = 5.1(4)$ Wm$^{-1}$K$^{-1}$, and an interfacial conductance of $h = 1.8(1)$ GWm$^{-2}$K$^{-1}$.

Without adding too much interpretation at this stage, we emphasize that the analysis shows a significant difference in the thermal transport properties of these two types of SLs. Namely, an increase of the volume conductivity by a factor of two, and a six-fold increase of the interfacial conductance can be seen by comparing SLs with m/n=2 with SLs with m/n=1. A first hint to where these differences are coming from can be obtained by estimating the volume conductivity $\kappa_v$ according to

$$\kappa_v^g = [m+n] / \left[\frac{m}{\kappa_{\text{LMO}}} + \frac{n}{\kappa_{\text{SMO}}}\right], \qquad (2)$$

which is the geometric average of the two contributing materials. In case of m/n=2, using the literature value of $\kappa_{\text{SMO}} = 4.5$ Wm$^{-1}$K$^{-1}$,[43] and the measured value of $\kappa_{\text{LMO}} = 1.3(1)$ Wm$^{-1}$K$^{-1}$, Eq.(2) yields $\kappa_v^g = 1.7(1)$ Wm$^{-1}$K$^{-1}$. This is a bit below, but still close to the $\kappa_v = 2.4(4)$ Wm$^{-1}$K$^{-1}$ derived from the linear fit of Eq. (1). For m/n=1, Eq. (2) yields $\kappa_v^g = 2.0(1)$ Wm$^{-1}$K$^{-1}$. This significantly underestimates the $\kappa_v = 5.1(4)$ Wm$^{-1}$K$^{-1}$ deduced from the experimental data. Therefore, the cause of the differences in $\kappa_v$ cannot simply be the different fractions of LMO and SMO. A possible origin is a change in the phonon spectrum, reflecting a change of the underlying crystal structure.

Regarding the strongly different Kapitsa conductances, we note here that XRD and XRR spectra reveal a small increase in the average RMS roughness of the LMO/SMO interfaces from $\sigma = 0.2(1)$ nm for $m/n = 1$ to $\sigma = 0.4(2)$ nm for $m/n = 1$ (see SI [37]). Since thermal conduction in atomic scale SLs is quite sensitive to roughness [2, 3], we can partially attribute the decrease in the interfacial conductance to this increase in roughness. Following the terminology of Swartz and Pohl [41], we here stress that, besides a change of the diffuse mismatch at the interface, also a change in the phonon spectral properties cause an increase in the acoustic mismatch, which can decrease the Kapitsa conductance.

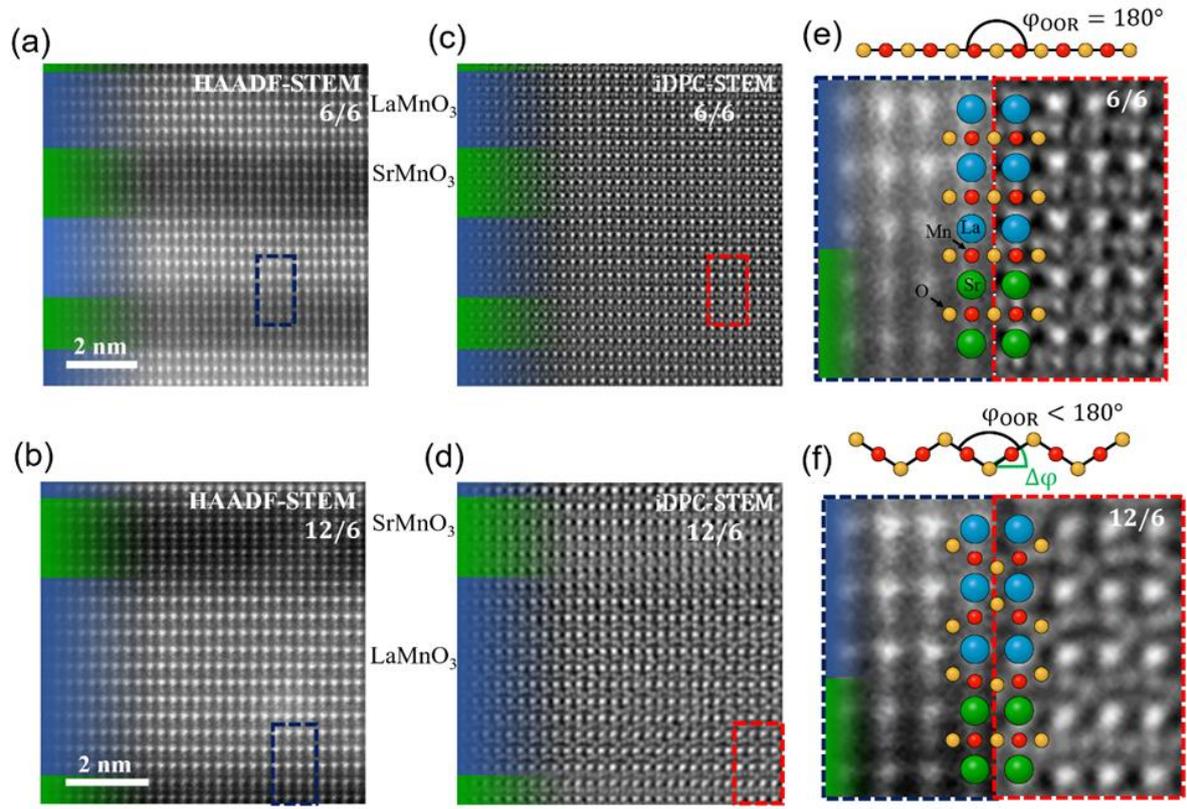

**FIG. 3.** Structural characterization of LMO/SMO SLs by TEM. Images were taken along the [110]-direction. (a) and (d) show HAADF-STEM image for SLs with m/n as indicated. (c) and (e) show corresponding iDPC-STEM images. (c) and (f) show enlarged images of the areas indicated in (a), (b), (d), (e) correspondingly, overlaid by a structural model.

## IV. STEM results

To clarify the origin of the puzzling changes in thermal properties, we have inspected our SLs by high-angle annular dark field scanning transmission electron microscopy (HAADF-STEM) and integrated Differential Phase Contrast (iDPC)-STEM [42, 43]. In Fig. 3(a) and (b), HAADF-STEM images of the SLs with m/n=1,2 demonstrate epitaxially grown layers with atomically sharp and flat interfaces with a roughness ≤ 1 unit cell (u.c.). From the iDPC-STEM images, shown in Fig. 3(c) and (d), we have estimated the Mn-O-Mn bond angle $\varphi_{OOR}$ in the LMO and SMO layers. For this purpose, we have quantitatively analysed contrast positions of oxygen and manganese columns, yielding the statistics of the bond angle deviation $\Delta\varphi=0.5(180°-\varphi_{OOR})$ with respect to the in-plane orientation (180°), within the image projection plane [110] (see Fig. 4). Note that, the aberration-corrected iDPC technique has been shown to be capable of pm precision when determining oxygen positions [44].

In detail, the Mn-O-Mn bond angles were determined for the 6/6 (Fig. 4(a)-(c)) and the 12/6 (Fig. 4(d)-(f)) SLs within the plane of zone axis projection, in terms of statistics of the bond angle deviation $\Delta\varphi=0.5(180°-\varphi_{OOR})$, resolved as a function of the atomic layer number, counted along the growth direction. Histograms, obtained from selected atomic layers within SMO and LMO are shown in Fig. 4(a), (d), and Fig. 4(b), (f), respectively. For a direct comparison, all histograms are plotted together in the colour-coded map depicted in the left panel of Fig. 4(c), and (f). From Gaussian fits we have determined the average angular deviation $\Delta\varphi_{av}$ and the corresponding variance $\sigma$ for each atomic layer. Results for 6/6 and 12/6 SLs are shown in the right panel of Fig. 4(c), (f).

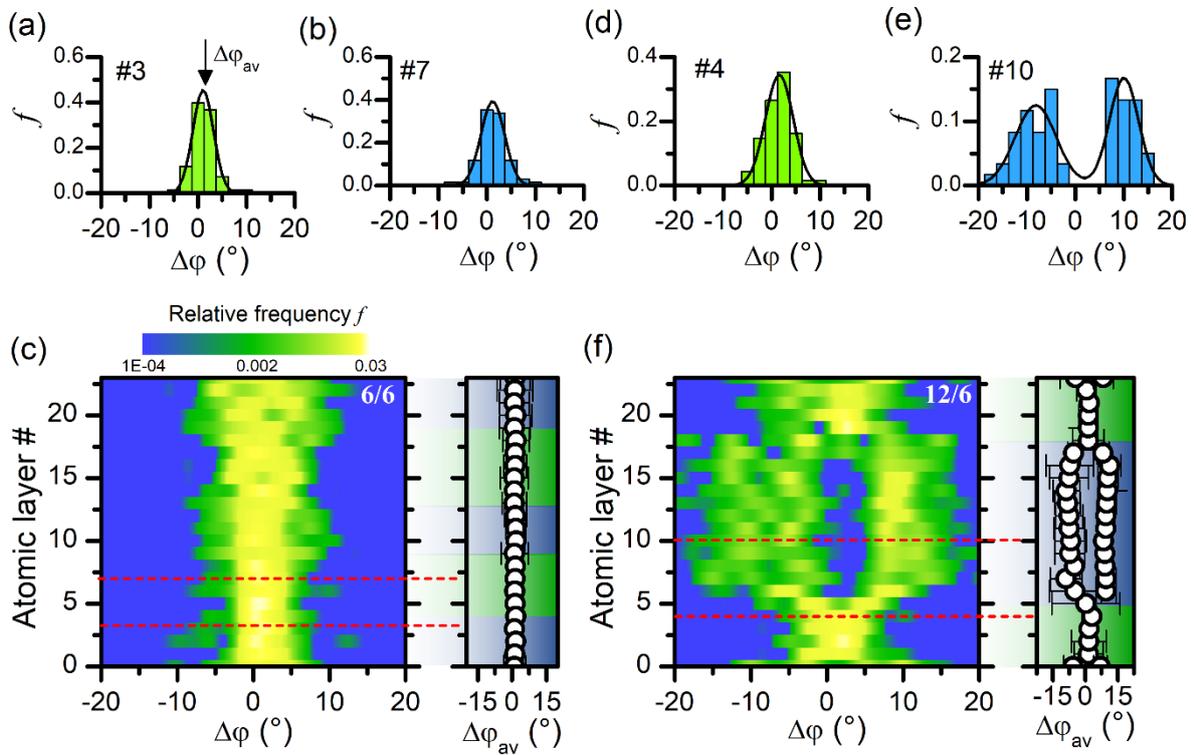

**FIG. 4.** Quantitative analysis of iDPC-STEM images in terms of angular deviation $\Delta\varphi$ of Mn-O-Mn rotation angle from 180°. Histogram (a) depicts the relative frequency of $\Delta\varphi$ for being within certain frequency intervals within a particular SMO layer as indicated inside SL with $m/n = 6/6$. The black curve is a Gaussian fit. (b) depicts a histogram for the same SL, but refers to a LMO layer. (c) The left panel shows all histograms interpolated along the angular axis as a color-coded logarithmic map. Dashed lines indicate the origin of the data shown in (a) and (b). The right panel shows the average angular deviation $\Delta\varphi_{av}$ for each atomic layer, determined by Gaussian fits. The error bar indicates the variance $\sigma$ of the distribution. In (d),

(e) and (f) similar plots as in (a), (b), (c) are shown, but here for a SL with $m/n = 12/6$. Note that in this SL a double Gaussian distribution within the LMO layer describes the data well.

This detailed analysis reveals that in all SLs the SMO layers possess $\varphi_{OOR} \approx 180°$, which is characteristic for bulk SMO [45]. In contrast, LMO layers inside the 12/6 SL display significantly smaller $\varphi_{OOR} < 180°$ as manifested by periodic zigzag-like features (see Fig. 3(d), (f)). The statistics obtained from the LMO layers in 12/6 SL reveals a bimodal distribution and $\varphi_{OOR} = 165° \pm 3°$, which is characteristic for bulk rhombohedral LMO [46,47]. Other m/n=2 SLs, i.e. 6/3 and 10/5 (see Fig. SM-5 in [37]) also show this behaviour. The situation changes in m/n=1 SLs, where both LMO and SMO layers possess the same $\varphi_{OOR} \approx 180°$. In Fig. 3(e) and (f) the enlarged images of selected areas in Fig. 3(a), (b), (c) and (d), overlaid by a structural model, highlight the presence and absence of octahedral distortion in the LMO layers in 12/6 and 6/6 SLs, respectively. SMO layers in the 12/6 SL display $\varphi_{OOR} = 180° \pm 3°$ similar to bulk SMO [45]. Thus, a large $MnO_6$ octahedral interfacial tilt mismatch is present in the m/n=2 SLs, whereas interfaces in the m/n=1 SLs are misfit free.

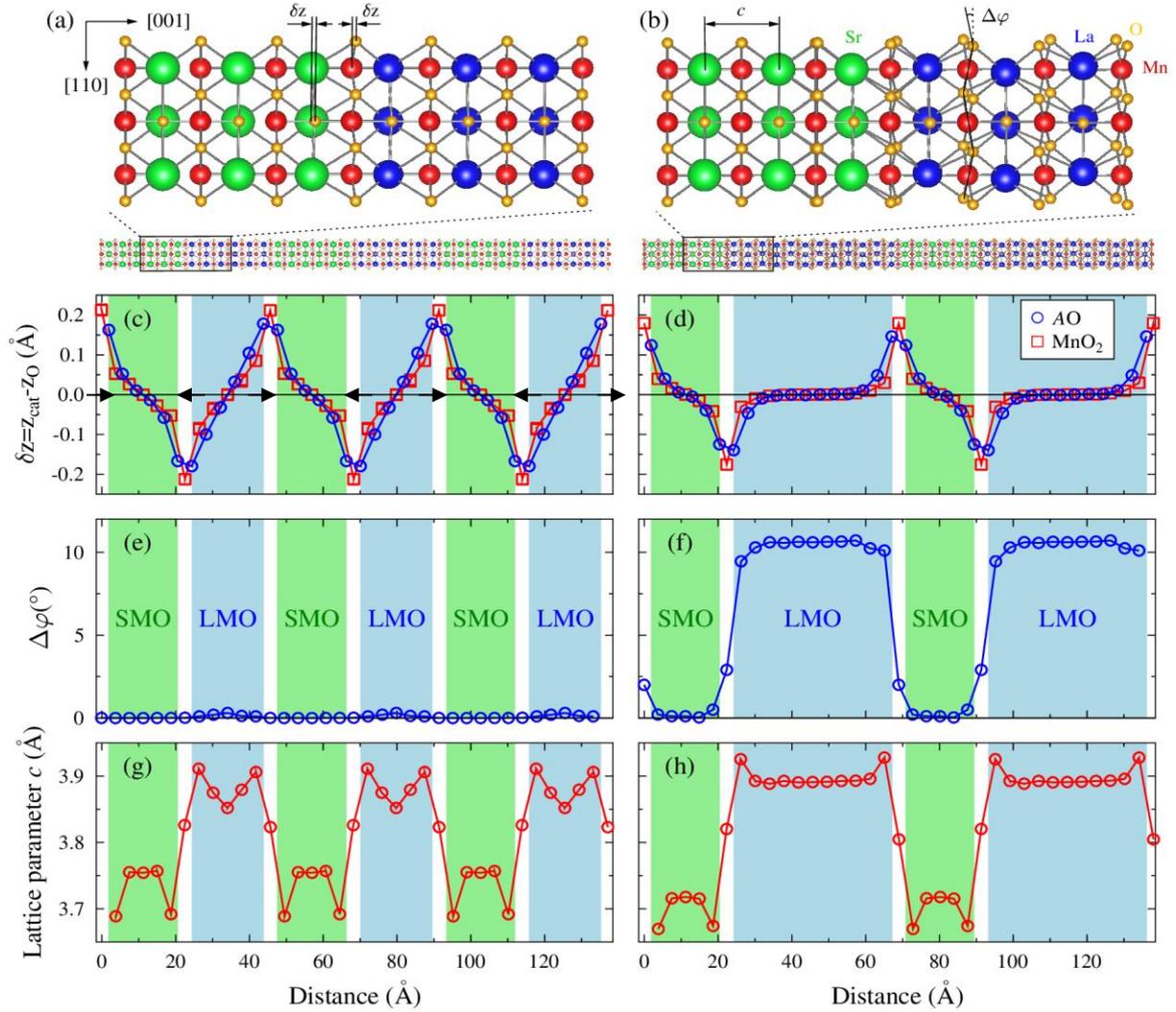

**FIG. 5.** Structure of the LMO$_m$/SMO$_n$ SLs optimized from first principles for m=6, 12 and n=6. The left panels (a), (c), and (e) show the details of 6/6 SL while the right panels (b), (d), and (f) illustrate the case of 12/6 sample. Atomic structure of m/n=1 is tetragonal distorted (a) along [001] axis, whereas in the case m/n=2 clearly the tilting of MnO$_6$ octahedra is clearly seen (b). The displacements of the cation/oxygen z-coordinates within the same layer, δz=z(cation)-z(O), as well as the bond angle deviations are plotted as a function of each atomic layer position along [001] in the panels (c) and (d), respectively. In the panel (c) the directions of electric dipole at the interface are marked by arrows. The out-of-plane lattice parameters, calculated as the La-La, La-Sr and Sr-Sr separations along [001], are plotted in the panels (e) and (f).

## V. DFT results

To disclose the origin of structural differences between the LMO and SMO layers in the m/n=1 and m/n=2 SLs we performed first principle simulations of the representative 6/6 and 12/6 SLs, using the √2x√2 in-plane geometry and constructing the supercells with 120 and 180 atoms, respectively. Then, the accurate structural optimization was performed using the code VASP [48], All LMO and SMO layers were assumed to have the same in-plane lattice parameter, a=0.3905 nm, due to epitaxy on the STO, while the coordinates of each atom were allowed to relax. The results presented in Fig. 5(a) and 5(b) demonstrate unambiguously the presence and absence of octahedral tilts, $\Delta\varphi$, in the LMO within the 12/6 and 6/6 SLs, respectively. For 6/3 SL, the simulations (see Fig. SM-6 in [37]) also reveal tilted LMO layers. The relaxed structure of the 6/6 and 12/6 SLs in Fig. 5(e) and 5(f) is tetragonally distorted along [001] as evidenced from the out-of-plane lattice parameters (the La-La, La-Sr and Sr-Sr distances). For both SLs their first interfacial SMO unit cells are always tetragonally compressed, while the scenario for LMO differs. For 12/6 SL the unit cell of LMO at the interface is tetragonally expanded compared to deeper lying layers. It seems that the LMO layers in 6/6 SL are not sufficiently thick to stabilize the tilted structure found in bulk samples at room temperature. However, for thicker LMO layers in 12/6 SL the tilting of the $MnO_6$ octahedra becomes energetically favourable as shown in Fig. 5(d) by the bond angle deviations $\Delta\varphi$ plotted for each atomic layer along the [001] axis. $\Delta\varphi$ for LMO layers vary robustly around 10°, except in the layers approaching the interfaces to SMO. There the boundary condition ($\Delta\varphi$=0) needs to be matched. Most importantly, the here calculated tilting angles $\Delta\varphi$ are in good agreement with STEM data in Fig. 3 and Fig. 4.

## VI. Correlation of spectral and structural properties

The HAADF-STEM, iDPC-STEM and DFT studies clearly show that the LMO layers in SLs with m/n=1 and in SLs with m/n=2 have a significantly different crystal structure, which is cubic for m/n=1 and rhombohedral for m/n=2. As pointed out before, a change in the crystal structure will change the phonon spectrum, and, therefore, will alter the thermal transport properties. Calculating such properties from first principles [49] is a tedious task for SLs. Fitting such models to experimental data is then often prohibitively expensive, such that one rather derives general trends and correlations from such modelling. To proceed, we here first-of-all present a survey over what is known about the phonon spectral properties of LMO and SMO.

For LMO, lattice dynamical calculations [50, 51], and neutron scattering from slightly doped La$_{0.8}$Sr$_{0.2}$MnO$_3$ [51] show that within the rhombohedral structural phase, the largest frequency of the acoustic phonons is about 3 THz. Also in the orthorhombic phase, neutron scattering and DFT calculation [52] reveal a similarly small acoustic phonon bandwidth. In contrast, lattice dynamical calculations [52, 54] of the cubic phase of LMO indicate a strong increase of the acoustic phonon bandwidth up to about 5.5 THz. Note that for the cubic phase LMO not much data can be found, because this structure is stable only at elevated temperatures above 750 K. For cubic SMO, DFT calculations [55], Raman [40] and neutron scattering [56] reveal a maximum acoustic phonon frequency of about 5 THz. Note that also in SMO, when the oxygen octahedra are rotated, as can be found in hexagonal SMO, the maximum acoustic frequency is reduced down to about 2 THz [55]. This overview suggests that the different thermal transport properties in our SLs is a consequence, which emerges from the fusion of two well matching phonon spectra (m/n=1) or two rather badly matching phonon spectra (m/n=2). The problem with this picture is that, in SLs with layers thicknesses of a few unit cells one can only speak about different phonon spectra if the phonon mean free path (mfp), $\xi$, is smaller than the individual layer thickness. If, instead $\xi > \Lambda$, the periodicity of the whole SL is directly encoded in the single, SL phonon spectrum.

### VII. Born-von-Kármán model

#### A. Description of the model

To assess the latter, we now introduce a Born-von-Kármán lattice dynamical model (BKM) [57]. More details of this model can be found in the appendix, and in the supplementary information [37]. The reason to introduce this model is not to provide an ab initio reproduction of the experimental findings, but to give a more quantitative version of the spectral matching argument sketched above and to estimate the phonon mfp. The primary task of the BKM is to deliver the acoustic phonon spectrum $\omega_i(\mathbf{k})$. Fig. 6 (a) shows a sketch of the model, including the modification which is incorporated to simulate rhombohedral (R$\bar{3}$c) LMO (see appendix B for details). With the spectrum $\omega_i(\mathbf{k})$ at hand, we estimate the cross-plane thermal conductivity by numerically evaluating

$$\kappa = \sum_i \int_{1.\text{BZ}} \frac{d^3\mathbf{k}}{(2\pi)^3} \hbar\omega_i(\mathbf{k}) v_z^i(\mathbf{k})^2 \tau_i(\mathbf{k}) \frac{\partial n(\omega_i, T)}{\partial T}, \qquad (3)$$

where $v_z^i(\mathbf{k}) = \frac{\partial \omega_i(\mathbf{k})}{\partial k_z}$ is the group velocity, and $\tau_i(\mathbf{k})$ is $\mathbf{k}$-dependent phonon life time, $n(\omega_i, T) = 1/\left(e^{\hbar\omega_i/k_B T} - 1\right)$ is the Bose-Einstein distribution, and the index $i$ refers to the phonon branch. Eq. (3) also provides a definition for a characteristic phonon mfp, projected onto the z-direction:

$$\xi_0 = \frac{\Sigma_i \int_{1.\mathrm{BZ}} \frac{d^3\mathbf{k}}{(2\pi)^3} \hbar\omega_i v_z^i(\mathbf{k})^2 \tau_i(\mathbf{k}) \frac{\partial n(\omega_i, T)}{\partial T}}{\Sigma_i \int_{1.\mathrm{BZ}} \frac{d^3\mathbf{k}}{(2\pi)^3} \hbar\omega_i v_z^i(\mathbf{k}) \frac{\partial n(\omega_i, T)}{\partial T}}. \qquad (4)$$

To evaluate Eqs. (3) and (4), we consider as scattering contributions according to Callaway [58] three-phonon Umklapp scattering $\tau_U(\omega, T) = A(T)\left(\frac{k_B}{\hbar\omega}\right)^2$, and Rayleigh scattering $\tau_R(\omega) = B\left(\frac{k_B}{\hbar\omega}\right)^4$. These contributions combine according to the Mathiessen rule to $\frac{1}{\tau} = \frac{1}{\tau_U} + \frac{1}{\tau_R}$.

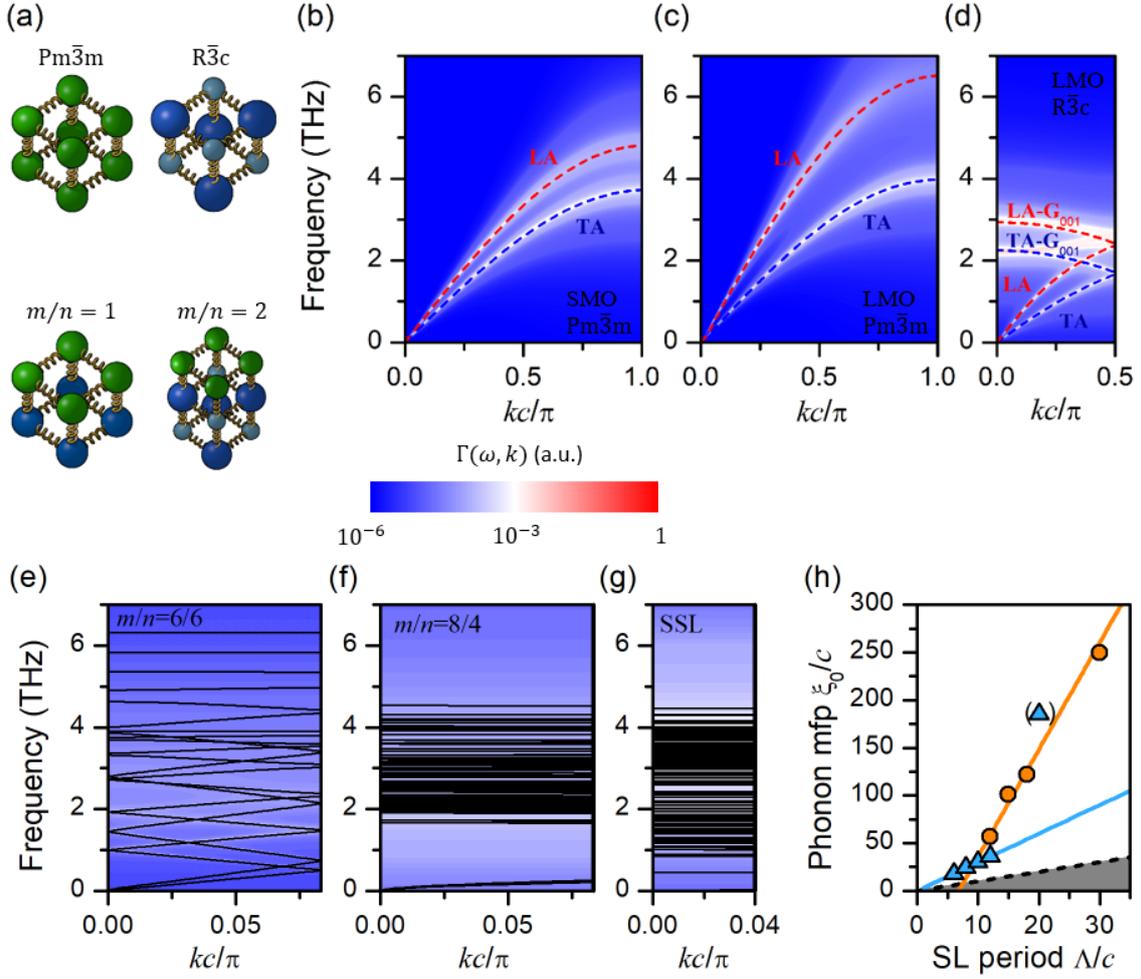

**FIG. 6.** Acoustic phonon spectra along $\Gamma X$ direction of pure manganite perovskites and SLs, calculated by the BKM. (a) Sketch of the model, visualizing effective atoms as spheres, and direct neighbour couplings as springs. (b) Spectrum of cubic SMO. Dashed lines are the calculated longitudinal (LA) and transversal (TA) phonon branches, the coloured map (log scale) in the background encodes the smearing of the phonon states due to finite life time according to Eq. (A6). (c) Spectrum of cubic LMO. (d) Spectrum of rhombohedral LMO. The origin of some of the back-folded branches is indicated by the related reciprocal lattice vector $G_{hkl}$. (e) SL with m/n=6/6, consisting of cubic LMO and cubic SMO. (f) SL with m/n=8/4, consisting of rhombohedral LMO and cubic SMO. (g) Phonon spectrum of the proposed phonon insulator with structure m/n/n/n=10/5/5/5. (h) Characteristic phonon mfp according to Eq. (4). Blue filled triangles refer to SLs with m/n=1, orange circles to m/n=2. Blue and orange solid lines are linear fits to these data. Note that the data point in brackets referring to the 10/10 SL was associated to the m/n=2 SLs. The black dashed line ($\xi_0 = \Lambda$) marks the transition between coherent and incoherent thermal transport.

## B. BKM results

Figure 6 summarizes the findings obtained from the BKM. In Fig. 6(b)-(d) acoustic phonon spectra of SMO (cubic) and LMO (cubic and rhombohedral) are shown. The coloured map in the background provides a measure for the relative importance of the phonon branches for thermal transport. These maps have been computed by assuming and a Lorentzian distribution in frequency-space for each mode (see Eq. (A6) in the appendix).

One can directly see that, the spectrum of cubic SMO matches well to that of cubic LMO. In contrast, for the rhombohedral LMO the reduced bandwidth implies a bad matching to cubic SMO. Details about choice of parameters for the life time can be found in the SI [37]. Figs. 6(e) and (f) show phonon spectra of SLs with m/n=6/6 and m/n=8/4. One can clearly see that, when stacking two cubic lattices, the SL spectrum folds back. However, the gaps opening up at $k = \frac{\pi}{\Lambda}$ are rather small. In contrast, a large gap of ~1.6 THz is introduced in the acoustic SL phonon spectrum in case of m/n=2. This gap restricts the three lowest bands to a maximum frequency of about 250 GHz. Note that, these bands provide the largest contribution to the thermal conductivity, as determined from Eq. (3). The modelling indicates that, a strongly reduced group velocity of these modes can explain the significant reduction of the volume thermal conductivity $\kappa_v$ in our SLs with m/n=2. Furthermore, fitting the measured thermal conductivities by the BKM via Eq. (3) allows to estimate the characteristic phonon mean free path according to Eq. (4). The result is shown in Fig. 6(h). For SLs with m/n=1 (blue triangles in Fig. 6(h)), the mfp turns out to be directly proportional to the SL period. A linear fit (blue line in Fig. 6(h)) yields $\xi_0 = 3\Lambda$. For SLs with m/n=2 (orange circles in Fig. 6(h)), a linear fit yields $\xi_0 = 11(\Lambda - 6c)$. Note that the SL with m/n=10/10 (data point in brackets in Fig. 6(h)) was modelled using a rhombohedral structure for LMO. Indeed, STEM investigations of the 10/10 SL shows OOR within the LMO layers similar to SLs with m/n=2 (see STEM images in Fig. SM-5 e) in ref. [37]). With this structure, the characteristic mfp complies well with the other SLs with m/n=2.

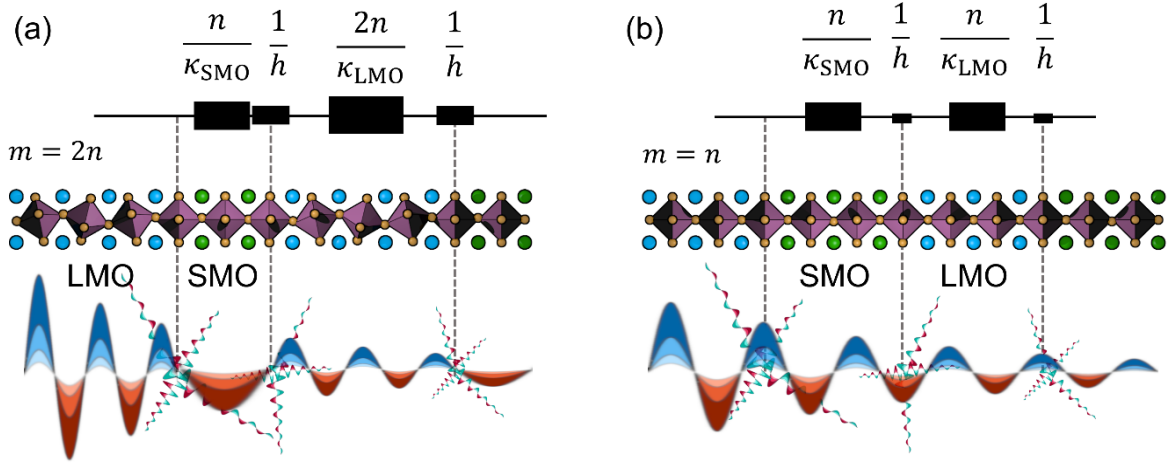

**FIG. 7.** Graphical summary of our investigations of thermal transport in LMO$_m$SMO$_n$ SLs with (a) m/n=2, and (b) m/n=1. In both subfigures, the top part shows the serial resistance model according to Eq. (1), applied in the analysis of the TTR experiments. The middle part depicts the structural model derived from STEM and DFT. The bottom part visualizes the final explanation for the difference in thermal properties based on SL phonon spectra derived from the BKM.

## VIII. Discussion

Figure 7 shows a graphical summary of our findings, which we discuss now in detail. In the beginning of this paper, we have reported TTR measurements of the thermal transport of (LMO)$_m$(SMO)$_n$ SLs. The analysis of the experimental data reveals strong differences in the thermal conductivity in SLs with different thickness ratio of LMO and SMO layers, m/n=1 and 2. The findings (see Fig. 2(d)) comply well with a simple serial resistance model (Eq. (1), top parts in Fig. 7). In particular, we were able to assign Kapitsa conductances $h$ ranging from 0.3 to 1.8 GWm$^{-2}$K$^{-1}$ to the LMO/SMO interfaces. The value of 1.8 GWm$^{-2}$K$^{-1}$ obtained in the octahedral-mismatch-free interfaces is surprisingly large in comparison with literature reports on other insulator-insulator interfaces [59], approaching the record values of 4 GWm$^{-2}$K$^{-1}$ measured in the metallic Al-Cu interface [60]. Advanced STEM imaging in combination with DFT calculations disclose a structural difference in the LMO inside the SLs. LMO is either cubic, that is without any oxygen octahedral rotation in SLs with m/n=1, or rhombohedral, displaying a significant OOR in SLs with m/n=2 (see Figs. 3, 4, 5 and the middle parts in Fig. 7). Note that, the fact that, large OOR correlates with small thermal conductivity among different manganites is empirically known since 1997 [61]. We here emphasize that, in this work we show that OOR and thus $\kappa$ can be structurally controlled within the same material by the architectural details of our atomic scale SLs.

Taking up the findings of STEM and DFT, the BKM shows that, due to the structural modifications of the LMO, the phonon spectra of the SLs are strongly different for the two kinds of SLs under consideration (see Fig. 6, and bottom parts in Fig. 7). This spectral difference in particular explains the difference of the volume thermal conductivity, which is the extrapolation of thermal conductivity to vanishing interface density. Going a step further, our estimation of the mfp shows a linear relation between mfp $\xi_0$ and SL period $\Lambda$ (see Fig. 6(h)). This implies that in all SL samples the dominant scattering processes giving rise to the large, but still finite $h$ occur in the interface region. This is indicated in the bottom sketches in Fig. 7 by the secondary phonons radiating away from the interfaces. In our SLs, Rayleigh scattering due to roughness can affect phonons with wave length $\lambda \approx 50$ u.c., given an interfacial waviness of this magnitude (see contrast modulation along interfaces in HAADF-STEM pictures shown in Fig. SM-5 in the SI [37]). Since these phonons represent only a small fraction of the full spectrum, increased lattice anharmonicity in the interfacial region may be equally important. As we discuss in the SI [37], an unambiguous discrimination of these two processes is not possible by means of the BKM, but the estimation of the mfp is quite insensitive on the assumptions made for the phonon life time entering Eq. (3) and (4).

We would like to emphasize that, in all samples $\xi_0 > \Lambda$, independent of the presence or absence of the OOR. This implies that, the thermal transport at room temperature is coherent with respect to the SL period. In SLs with m/n=1, the linear extrapolation of the mfp to $\Lambda = 2c$ (see Fig. 6(h)) shows that even in the atomic limit, ballistic transport can be expected. In contrast, in SLs with m/n=2, extrapolating the linear fit shown in Fig. 6(h) predicts a transition to incoherent transport at $\Lambda = 6$. Note that this number coincides with twice the width of the transition zone from cubic to rhombohedral found in the statistical analysis of the iDPC-STEM pictures (see Fig. 4(f)) and in the DFT simulations (see Fig. 5(f)). Thus, we conclude that this effective broadening of the interface likely prevents ballistic transport in the truly atomic SL limit in case of m/n=2.

As an outlook, we would like to briefly discuss a suggestion for an SL architecture, which leads to a phonon insulator. The unit cell of this thermal metamaterial starts with a thick (m≥10), and thus rhombohedral LMO layer, followed by three thin (for each layer n=m/2), and thus fully cubic layers of SMO, LMO, and SMO. The structural formula of this super-superlattice (SSL) is $(LMO)_{10}(SMO)_5(LMO)_5(SMO)_5$ (SSL with m/n/n/n=10/5/5/5). The BKM of this SSL structure yields the phonon spectrum shown in Fig. 6(g). Note that, the three lowest phonon branches are now so strongly reduced in frequency, that they cannot even be seen in this diagram. As a consequence, the group velocities decrease such that the characteristic mfp of $\xi_0 = 11.4c$ drops below one unit cell of the SSL. This means that each mode is strictly confined within certain parts of the SSL only. Such local resonances cannot transport energy, and thus one can expect that the thermal conductivity of the SSL vanishes. Cohn et al. [61] estimated for rhombohedral LMO at room temperature in the so-called amorphous limit [62] a thermal conductivity of about 0.7 Wm$^{-1}$K$^{-1}$. The SLs with m/n=2 are exactly in this range. Estimating $\kappa$ from Eq. (3) for the proposed SSL yields a value below 0.01 Wm$^{-1}$K$^{-1}$. This implies that, an SSL can beat the amorphous limit by more than one of magnitude.

## IX. Summary

In summary, combining transient thermal reflectivity, advanced STEM imaging methods, ab density functional theory, and a classical lattice dynamical model, we have shown a significant impact of structural modifications in LMO/SMO SLs on their thermal transport properties. Recall that, in the introduction, we have emphasized the critical role of thermal materials research for efficiency of energy conversion processes. Our findings suggest to exploit non-equilibrium structural phases (here, the cubic phase of LMO), which only emerge within a phononic crystal, as a design element for novel thermal metamaterials.

This work was funded by the Deutsche Forschungsgemeinschaft (DFG, German Research Foundation) - 217133147/SFB 1073, projects A02, Z02. I.V.M. and S.O. acknowledge finding by the European Union (EFRE).

## X. Appendix

### A. Further details of the BKM

In this appendix we explain the BKM in more detail. In this classical, lattice dynamical model, we consider only acoustic phonons. To understand this approximation, recall that, at room temperature optical phonons in manganites have frequencies $\omega \geq \omega_{\text{th}}$. Typically, the life times for these modes are very short, and the group velocities are often small compared to

$v_{LA}$ and $v_{TA}$. As a consequence, they contribute little to the thermal conductivity according to Eq. (3), and, therefore, we neglect them in the BKM. Within the framework of the latter, we consider lattices hosting effective atoms with masses $m_{LMO} = m_{La} + m_{Mn} + 3m_O$ and $m_{SMO} = m_{Sr} + m_{Mn} + 3m_O$. These atoms interact in harmonic approximation with their direct, next, and next-next neighbours. This gives rise to the following general equation of motion for the displacement components $u_i$ ($i$=x,y,z) of the $k$-th effective atom inside the unit cell at $\mathbf{R_{mno}}$:

$$m_k \frac{\partial^2 u_i(\mathbf{R_{mno}},t)}{\partial t^2} = -\sum_\alpha K_\alpha \sum_{j=1...j_\alpha} \hat{n}_{ji} \left[\mathbf{u}(\mathbf{R_{mno}},t) - \mathbf{u}(\mathbf{R_{mno}}+\mathbf{n}_j,t)\right] \cdot \hat{\mathbf{n}}_j, \quad (A1)$$

where $\alpha$ refers to direct, next, and next-next neighbours, $j_\alpha$ refers to the number of latter (6,12, and 8), $\mathbf{n}_j$ is the vector pointing to a specific neighbouring atom, and $\hat{\mathbf{n}}_j = \mathbf{n}_j/|\mathbf{n}_j|$. The $K_\alpha$'s are the harmonic force constants, which parametrize the interaction strength towards the neighbouring atoms. From the usual ansatz of plane waves, that is $\mathbf{u}(\mathbf{k},\mathbf{t}) \sim e^{i\mathbf{kx}-i\omega t}$, one obtains the phonon spectrum $\omega(\mathbf{k})$ by solving the eigenvalue problem

$$\overline{D}(\mathbf{k}) \cdot \mathbf{e} = \omega^2 \mathbf{e}. \quad (A2)$$

Here, $\overline{D}(\mathbf{k})$ is the dynamical matrix arising from compiling the $3k$ equations defined by Eq. (5), and $\mathbf{e}$ encodes the polarization of the phonon eigenmodes. Note that, compared to the model employed in the seminal work of Simkin and Mahan [1], our BKM distinguishes between longitudinal and transversal modes.

To calculate the thermal conductivity from the spectrum according to Eq. (3), the first Brillouin zone is systematically rasterized with a resolution of $\Delta k_x = \Delta k_y = \pi/a/N$, and $\Delta k_z = \pi/\Lambda/N$, and at each point in k-space the spectrum is computed by solving Eq.(6). For SLs with m/n=1 (m/n=2) we have used N=100 (N=50). We checked that, a finer resolution in reciprocal space does not alter the resulting thermal conductivity. The group velocities entering Eq. (3) were calculated from a finite difference approximation according to

$$v_z(\mathbf{k}) = \frac{\omega(k_x,k_x,k_x+\Delta k_z) - \omega(k_x,k_x,k_x-\Delta k_z)}{2\Delta k_z}. \quad (A3)$$

As a simplified alternative to Eq. (3) following [1], one may consider a constant phonon mean free path $\xi$ for all modes at all frequencies. Then, one finds

$$\kappa = \xi \sum_i \int_{1.BZ} \frac{d^3\mathbf{k}}{(2\pi)^3} \hbar\omega_i(\mathbf{k}+i/\xi) v_z^i(\mathbf{k}+i/\xi) \frac{\partial n(\omega_i,T)}{\partial T}. \quad (A4)$$

Furthermore, we define in analogy to Eq. (4) the characteristic frequency

$$\omega_0 = \frac{\sum_i \int_{1.BZ} \frac{d^3\mathbf{k}}{(2\pi)^3} \hbar\omega_i v_z^i(\mathbf{k})^2 \tau_i(\mathbf{k}) \frac{\partial n(\omega_i,T)}{\partial T}}{\sum_i \int_{1.BZ} \frac{d^3\mathbf{k}}{(2\pi)^3} \hbar v_z^i(\mathbf{k}) \tau_i(\mathbf{k}) \frac{\partial n(\omega_i,T)}{\partial T}}. \quad (A5)$$

The coloured maps $\Gamma(\omega, \mathbf{k})$ in the background in Fig. 6 (b) to (g) were calculated by assuming a Lorentzian function for each phonon mode, that is

$$\Gamma(\omega, \mathbf{k}) = \sum_i \frac{0.5\omega_r}{(\omega-\omega_i(\mathbf{k}))+(0.5\omega_r)^2}, \quad (A6)$$

where $\omega_r = \frac{1}{\tau_U} + \frac{1}{\tau_R}$.

### B. Modifications for rhombohedral LMO

For simulating rhombohedral LMO, we consider 2 by 2 by 2 atoms in an enlarged unit cell with 8 effective atoms. The masses are varied in a checkerboard-like fashion with $m_{LMO}^1 = (2 - 10^{-5}) m_{LMO}$ and $m_{LMO}^2 = 10^{-5} \cdot m_{LMO}$ (represented by large dark blue, and smaller light blue spheres in Fig. 6(a)). This approach yields in addition to a realistic acoustic phonon part in the spectrum, a set of unrealistic high frequency phonon branches, several 100 THz above the TA and LA modes. These phonon modes reside solely on the small-mass atoms. Being an unphysical modelling artefact, we ignore these phonon branches. This reduces the total number of modes by a factor of two. In a more realistic modelling approach [55], these neglected phonon branches are part of the optical phonon spectrum directly above the acoustic phonons, which are considered in our approach exclusively, as justified above. We here emphasize that, with our modelling ansatz for rhombohedral LMO, the correct mass density is maintained. We reiterate at this point that, the purpose of the BKM is not to provide an ab initio way to reproduce our experimental findings, but rather to provide a more intuitive understanding of the thermal transport properties in the different SL architectures. The particular choices for harmonic force constants $K_\alpha$ and the life time parameters $A$ and $B$ are further described and justified in the Supplementary Information [36].

# Atomic scale spectral control of thermal transport in phononic crystal superlattices

## Supplementary Information

D. Meyer[1], V. Roddatis[2,3], J.P. Bange[1], S. Lopatin[4], M. Keunecke[1], D. Metternich[1], U. Roß[3], I.V. Maznichenko[5], S. Ostanin[5], I. Mertig[5], V. Radisch[3], R. Egoavil[6], I. Lazić[6], V. Moshnyaga[1*], and H. Ulrichs[1**]

[1] *Erstes Physikalisches Institut, Georg-August-Universität Göttingen, Friedrich-Hund-Platz 1, 37077 Göttingen, Germany*
[2] *GFZ German Research Centre for Geosciences, Helmholtz Centre Potsdam, Telegrafenberg, 14473 Potsdam, Germany*
[3] *Institut für Materialphysik, Georg-August-Universität Göttingen, Friedrich-Hund-Platz 1, 37077 Göttingen, Germany*
[4] *Core Lab King Abdullah University of Science and Technology, Thuwal 23955, Saudi Arabia*
[5] *Institut für Physik, Martin-Luther-Universität Halle-Wittenberg, D-06120 Halle, Germany*
[6] *Thermo Fisher Scientific (formerly FEI), Achtseweg Noord 5, 5600KA, Eindhoven, The Netherlands*

*e-mail of corresponding author 1: vmosnea@gwdg.de

**e-mail of corresponding author 2: hulrich@gwdg.de

In this supplement, we provide details about A) the sample growth method, B) the structure of the samples, C) the TTR setup, D) the thermal three-layer model, E) TEM characterization, F) DFT calculation, and G) about the lattice dynamical model.

## A) Sample Growth

The $LMO_m/SMO_n$ SLs were grown on $SrTiO_3(100)$ substrates by a metalorganic aerosol deposition (MAD) technique by using La(III)-, Sr(II)- and Mn(II) acetylacetonate (acac) as precursors. The LMO and SMO precursor solutions in dymethilformamide with empirically determined molar ratios $La(acac)_3/Mn(acac)_2=1.3$ and $Sr(acac)_2/Mn(acac)_2=1.2$ were alternatingly sprayed by using of dry compressed air onto the substrate heated up to 900°C at ambient atmospheric pressure conditions, i.e. $pO_2=0.2$ bar. The volumes of precursor solutions, required for the deposition of one monolayer of LMO and SMO, were determined from the growth of single LMO and SMO films with thicknesses, d=5-20 nm. The growth of SLs was monitored in situ by optical ellipsometry analogously to that reported in Ref. [34] as exemplarily shown in Fig. SM-1 for 10/5 and 5/5 SLs.

## B) Structural Characterization

Global structural characterization was performed by X-ray diffraction (XRD) and X-ray

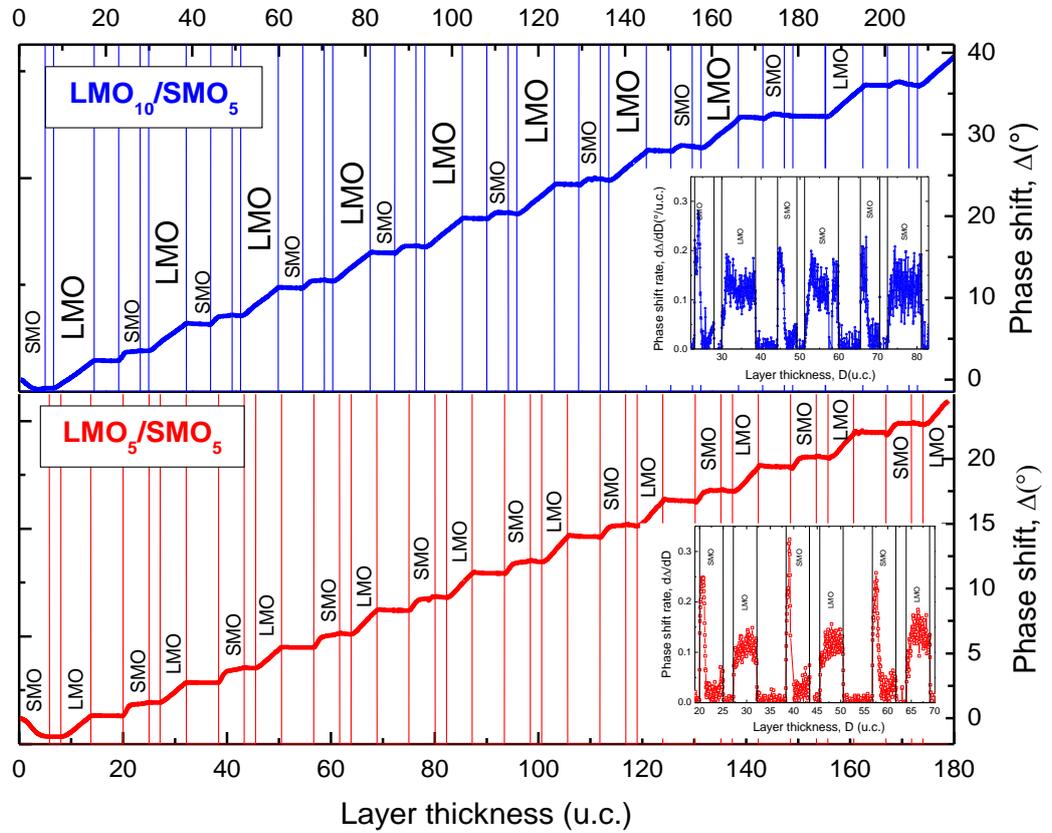

**Figure SM-1:** Development of the phase shift angle, $\Delta$, as a function of layer thickness, $D$, monitored in situ during the growth of $LMO_{10}/SMO_5$ (top panel, blue curve) and $LMO_5/SMO_5$ (bottom, red) SLs. The measured $\Delta(t)$ diagrams were renormalized to $\Delta(D=v*t)$ dependences, using the known constant deposition rate $v=0.3$ u.c./s, independently verified by X-ray reflection and TEM data. The vertical lines marked the opening and closing of LMO and SMO liquid channels. One can see an almost linear $\Delta(D)$ behavior for LMO layers and a non-monotonous behavior for SMO layers due to the charge transfer from the underlying LMO into the SMO layer (see also ref. 1). The insets show the zoomed view of the phase shift rate, $d\Delta/dD$, as a function of layer thickness, which, being a measure of charge density within the layer (Ref. [34]), illustrates the in situ charge transfer across the interface from the LMO into the SMO and the increased electron concentration within the first 1-2 u.c. of SMO layers close to the SMO(top)/LMO(bottom) interface.

reflection (XRR) using a using Bruker D8 diffractometer with Cuk$_\alpha$ radiation. XRR and XRD patterns were simulated by programs packages described in Ref.s [35, 36]. For the SL with

m/n=10/5 (see Fig. 1(a) and (b)), fitting the XRR spectrum yields RMS roughness's of $\sigma = 1.1(1)$ nm (STO/SL), $\sigma = 0.3(2)$ nm (LMO/SMO), $\sigma = 0.5(2)$ nm (SMO/LMO), and $\sigma = 0.4(1)$ nm (SL surface). Fitting the XRD spectrum of the m/n=10/5 SL yields RMS roughness's of $\sigma = 1.0(5)$ nm (STO/SL), $\sigma = 0.3(2)$ nm (LMO/SMO), $\sigma = 0.6(3)$ nm (SMO/LMO), and $\sigma = 0.6(4)$ nm (SL surface). For the SL with m/n=5/5, fitting the XRR spectrum yields RMS roughness's of $\sigma = 1.0(3)$ nm (STO/SL), $\sigma = 0.3(1)$ nm (LMO/SMO), $\sigma = 0.3(1)$ nm (SMO/LMO), and $\sigma = 0.3(1)$ nm (SL surface). Fitting the XRD spectrum of the m/n=5/5 SL yields RMS roughness's of $\sigma = 1.0(5)$ nm (STO/SL), $\sigma = 0.2(1)$ nm (LMO/SMO), $\sigma = 0.4(3)$ nm (SMO/LMO), and $\sigma = 0.5(3)$ nm (SL surface). In Table SM-I we summarize further structural parameters derived from XRR and XRD. The surface morphology of SLs was inspected at room temperature using an atomic force microscope (AFM) from Innova-Bruker.

| | XRR | | | XRD | |
|---|---|---|---|---|---|
| $m/n$ | $d_{SMO}$ (nm) | $d_{LMO}$ (nm) | $\Lambda$ (nm) | $c_{oop}$ (nm) | $m_{exp}/n_{exp}$ |
| 20/10 | 3.8 | 8 | 11.6 | 0.385 | 20.5/10 |
| 12/6 | 2.1 | 2.2 | 6.7 | 0.385 | 11.2/6.1 |
| 10/5 | 2.0 | 3.9 | 5.9 | 0.384 | 10.1/5.4 |
| 8/4 | 1.4 | 3.0 | 4.5 | 0.384 | 7.8/3.8 |
| 6/3 | 1.2 | 2.2 | 3.5 | 0.385 | 5.6/3.2 |
| 10/10 | 3.5 | 3.9 | 7.2 | 0.382 | 10/9.2 |
| 6/6 | 2.0 | 2.2 | 4.2 | 0.382 | 5.6/5.5 |
| 5/5 | 1.7 | 2.0 | 3.8 | 0.382 | 5.3/4.6 |
| 4/4 | 1.5 | 1.5 | 3.0 | 0.382 | 3.9/4 |
| 3/3 | 1.1 | 1.1 | 2.3 | 0.382 | 2.9/2.9 |

**Table SM-I:** Structural properties obtained from X-ray reflectivity (XRR) and X-ray diffraction (XRD) for all SLs. Individual thicknesses of layers $d_{SMO}$ and $d_{LMO}$ were obtained from XRR simulations by using of ReMagX [35], and the mean out-of-plane lattice parameter $c$ from XRD simulations [36]. $n_{exp}$ and $m_{exp}$ are evaluated from thicknesses of SMO and LMO layers, and $c$, respectively.

## C) Transient Thermoreflectance Setup

Transient thermoreflectance (TTR) is a contactless optical pump-probe method to determine the thermal conductivity, κ, interfacial thermal conductance $h$ and the specific heat, $C_P$, of thin film samples [38, 63, 64]. As the complex refractive index $N = n + ik$ is typically temperature dependent, the reflectivity, $R = [(n-1)^2 + k^2]/[(n+1)^2 + k^2]$, of a material depends on temperature. The temperature dependent change in reflectivity can be expressed in a first order Taylor expansion as $\Delta R/R = C\Delta T_{\text{surf}}$ (reference [11]). We call $C$ the thermoreflection coefficient, which is usually in the range of $10^{-2}$ K$^{-1}$ to $10^{-5}$ K$^{-1}$ (Ref. [38]). Note that, $C$ typically depends on the wavelength of the probing light and on the angle of incidence

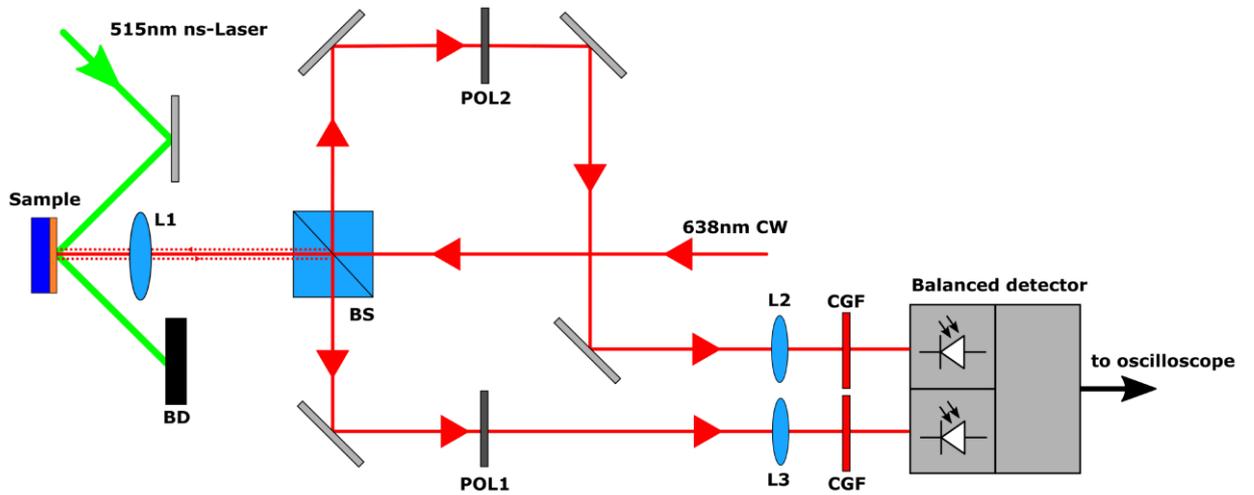

**Figure SM-2:** Detailed sketch of our TTR setup.

For the thermal conductivity measurements discussed in this article we employ a custom-built optical setup (see Fig. SM-8). A pulsed pump laser (Coherent Inc., FLARE NX, wavelength λ=515 nm, pulse duration 1 ns, repetition rate 2 kHz) is used to rapidly heat the sample surface ($\Delta T_{\max} \approx 100$ K) covered by an absorber film (Cu with a thickness of 50 nm). The energy of the pump beam is reduced to 90 mJ using a 0.5 neutral density filter. The beam diameter of the unfocused pump beam on the sample surface is 1.9 mm. As a probe laser, we use a diode continuous-wave laser with a wavelength of 643 nm, and an adjustable output power with maximum 150 mW (TOPTICA Photonics AG, iBEAM-SMART-640-S). The beam diameter of the focused probe beam on the sample surface is 23 mm.
For the measurements reported in this article the power was set to 50 mW. The temporal evolution of the surface temperature is monitored by a sampling oscilloscope (Agilent Technologies DSO-X 3054A, 1 GHz band width), detecting via a fast, balanced photodiode

(FEMTO Messtechnik GmbH, HCA-S) the change in optical reflectivity $\Delta R \propto \Delta T_{surf}$ of the probe laser directly in the time domain. The oscilloscope is triggered by a reference photo detector (EOM) observing the pump pulses.

The actual beam paths of the lasers, as well as all optical components are sketched in Fig. SM-8. The probe beam passes a 50:50 cube beam splitter **BS**, and is thereby separated in a probe and a reference beam. The probe beam is collinearly focused by the lens **L1** onto the samples surface (spot diameter 23 µm). The reflected beam is again collected by **L1**, and split from the incoming beam by passing the beam splitter **BS**. After passing the thin-flm polarizer **POL1**, a lens **L3** focuses the beam into a balanced photo detector. The reference beam, splitted at **BS** is likewise passing a polarizer **POL2**, and focused with lens **L2** on the reference port of the photo detector. The polarizers can be used to manually balance the power in the probe and the reference beams. Two colored glass filters **CGF** directly in front of the balanced photo detector filter unwanted pump light from the probe and reference beams. The pump beam is directed at the sample at an angle of 45°. The reflection from the sample is dumped into a block of anodized aluminium **BD**. An off-axis CCD-camera is used to monitor position and overlap of the beams on the sample, and to check for laser induced damage. Typical raw measurement data is shown in Fig. SM-3. The samples are mounted inside an evacuated cryostat with optical access. Nitrogen cooling and resistive heating allows to control the base temperature $T_0$ between 100 K and 400 K.

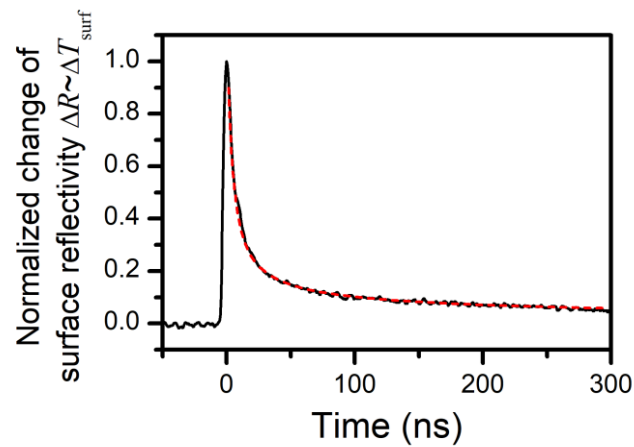

**Figure SM-3**: A typical TTR curve, obtained from the SL with m/n=5/5 (continuous black line) at 293 K, together with a fit of our thermal model Eq. (S8) (red dashed line).

While theoretically all materials with a sufficiently high reflectivity can be probed this way, the signal-to-noise ratio can be drastically increased by using a thermal transducer layer with a high thermo-reflectance coefficient $C$, and good absorption. For our probe light (643 nm) copper, which has a rather large $C_{Cu} = 8.5*10^{-5} K^{-1}$ (Ref. [65]) is a reasonable choice. Due to the large thermal conductivity of copper, this metallic material also simplifies the subsequent analysis: since the time resolution of the oscilloscope is as short as the pump pulse (1 ns), and thermal equilibration in the copper layer is even faster (a few picoseconds), the initial temperature increase can be considered as being instantaneously spatially homogeneous inside the copper layer. This eliminates the need for modeling the lasers energy transfer as a function of the absorption depth. For these reasons, all SL samples have been covered by a 50 nm thick copper film grown by means of electron beam physical vapor deposition in UHV conditions. The film thickness was controlled in-situ by a quartz crystal balance, and cross-checked ex-situ by XRR measurements.

**D) Thermal three-layer model**

To extract information about the thermal conductivity of the SL samples, the measured TTR curves are compared to an analytic description of thermal transport, based on a modelling technique introduced by Balageas et al. [39]. Within this model, the SL is considered as a single film with homogenous material parameter. Then, in all parts of the samples, classical diffusion equations can be used to describe thermal transport. Due to the large difference in pump and probe spot diameters, we can safely assume the direction of the heat transport to be purely normal to the sample surface. The mathematical problem then reduces to a single spatial dimension. The basic system of equations to be solved consists of classical Fourier diffusion equations

$$\frac{\partial \theta_i}{\partial t} = \frac{\kappa_i}{\rho_i c_i} \frac{\partial^2 \theta_i}{\partial z^2}, \qquad (S1)$$

where $\kappa_i$ is the thermal conductivity, $\rho_i$ is the density, and $c_i$ is the specific heat of layer $i$ ($i = 1$ refers to the Cu layer, $i = 1$ to the SL, $i = 3$ to the substrate), and $\theta_i = T_i - T_0$ is the temperature excursion with respect to the base temperature $T_0$. As boundary conditions, we consider Kapitza interface conditions between adjacent layers ($z = z_i$):

$$-\kappa_i \frac{\partial \theta_i(z_i,t)}{\partial z} = -\kappa_{i+1} \frac{\partial \theta_{i+1}(z_i,t)}{\partial z} = h[\theta_i(z_i,t) - \theta_{i+1}(z_i,t)], \qquad (S2)$$

where $h$ is the interfacial thermal conductance. The pump laser enters into the boundary and initial condition at the surface of the sample

$$-\kappa_1 \frac{\partial \theta_1(0,t)}{\partial z} = \delta(t). \qquad (S3)$$

The bottom of the substrate ($z = z_b = t_{Cu} + t_{SL} + t_{sub}$) is fixed to the base temperature. This leads to

$$-\kappa_3 \frac{\partial \theta_3(z_\mathrm{b},t)}{\partial z} = 0. \qquad (S4)$$

Upon applying a Laplace transform, the time-derivatives vanish and become multiplicative expressions. The remaining normal differential equations can be solved with the following ansatz for each layer:

$$\tilde{\theta}_i(z,s) = A_i(s) \sinh\left((z - z_{i-1})\sqrt{\frac{\rho_i c_i}{\kappa_i}s}\right) + B_i(s) \cosh\left((z - z_{i-1})\sqrt{\frac{\rho_i c_i}{\kappa_i}s}\right). \qquad (S5)$$

This results in a system of six linear equations for the coefficients $A_i(s), B_i(s)$. In a compact notation, one can write

$$\overline{\mathrm{M}} \cdot \mathbf{a} = (1,0,0,0,0,0)^\mathrm{T}. \qquad (S6)$$

Here, **a** consists of the unknown coefficients $(A_i(s), B_i(s))$ in the ansatz (S5). The system of equations (S6) can be solved for the surface temperature excursion in Laplace-space, which corresponds to the second entry of $\vec{a}$:

$$\tilde{\theta}_\mathrm{surf}(s) \sim 1 + \sum_{i=1}^{\infty} \frac{\det \overline{\mathrm{M}}_{12}^{\downarrow}}{\det \overline{\mathrm{M}}}, \qquad (S7)$$

where $\overline{\mathrm{M}}_{12}^{\downarrow}$ is the minor matrix, which is obtained by omitting the first row and second column of $\overline{\mathrm{M}}$. To obtain the time dependent expression, the inverse Laplace transform of $\tilde{\theta}_\mathrm{surf}(s)$ has to be calculated by inverse Laplace transformation. The residual theorem yields:

$$\theta_\mathrm{surf}(t) \sim 1 + \sum_{i=1}^{\infty} \frac{\det \overline{\mathrm{M}}_{12}^{\downarrow}}{\frac{\partial \det \overline{\mathrm{M}}}{\partial s}} e^{s_i t}. \qquad (S8)$$

Note that, in (S8), the real, negative roots $s_i$ of $\det \mathrm{M}$ enter into the argument of the exponential function. These roots have to be determined numerically. In practice, the sum in (S8) is truncated after the first 2000 terms. Larger summands only contribute to very short time scales < 5 ns directly after the excitation. To implement a better approximation of the true pulse shape, we apply the modification of (S8) described in detail in [39].

|  | $c_p$ (J kg$^{-1}$ K$^{-1}$) | $\kappa$ (Wm$^{-1}$K$^{-1}$) | $\rho$ (kg m$^{-3}$) | $h_{ij}$ (GWm$^{-2}$K$^{-1}$) |
|---|---|---|---|---|
| Cu | 385 [38] | 400 [63] | 8960 |  |
| SL (*m/n=2*) | 511 | * | 6443 | 0.009(1) (Cu/SL) |
| SL (*m/n=1*) | 532 | * | 6222 | 0.087(1) (Cu/SL) |
| STO | 544 [64] | 12 [64] | 5110 | 20(1) (SL/STO) |

**Table SM-II:** This table shows assumed values for heat capacities $c_p$, thermal conductivities κ and mass densities at 293 K, mass densities $\rho$, and the interface conductances $h_{ij}$ resulting from the fit of the thermal model to the experimental TTR curves. *See Figure 2, and Table SM-IV.

To fit the thermal model to the experimental data, we use non-linear optimization in the form of a least squares-fit. Free parameters are the interface conductivities between the SL and the Cu layer, and between the SL and the substrate, as well as the effective thermal conductivity of the SL. The other necessary parameters for the description of the samples are taken from literature [38, 63 – 72], and are listed in Table SM-II (for temperature dependencies see Fig. SM-4). Note that, we assume that, the interface conductivities do not depend on the SL period. Thus, we simultaneously optimize $h_{12}$ and $h_{23}$ as shared parameters for all samples with the same *m/n*, while the $\kappa_{\text{SL}}$'s are optimized individually within this fitting procedure. The optimization method of choice is the Nelder-Mead method, which is a derivative free algorithm and therefore well-suited for the rather involved evaluation of the thermal model. It optimizes the parameters by iteration, where in each step the values on the corners of a multidimensional simplex are compared and said simplex modified (e.g. by scaling, mirroring or stretching). For each sample, the thickness of the SL layer taken from the X-Ray measurements summarized in Table SM-I. The heat capacities $c_p$ at 293 K and densities of the SLs were interpolated from material parameters of LSMO [66 – 68], according to the actual La and Sr content for a given *m* and *n*. When considering base temperature other than 293 K, the temperature dependencies depicted in Figure SM-4 were used. Note that, in the generic temperature dependency of $c_p(T_0)$ used for the SLs local maxima due to phase transition are not included. This can be justified by fact that the molecular heat capacities of many different manganites are approximately identical, regardless of the actual structure (R-3c, Pm-3m etc.). This is likely related to the dominance of optical phonons in $c_p$. Nevertheless, such an interpolation for $c_p$ introduces potential errors in the determination of $\kappa_{\text{SL}}$. Varying $c_p$ within a range of 10 % around the assumed value at 293 K shows that, this error in $\kappa_{\text{SL}}$ is about 1.4%, which is smaller than the typical mean deviation resulting from determining $\kappa_{SL}$ from three independent measurements. The mass densities of the SLs listed in Table SM-II were calculated from the lattice parameters determined by XRD.

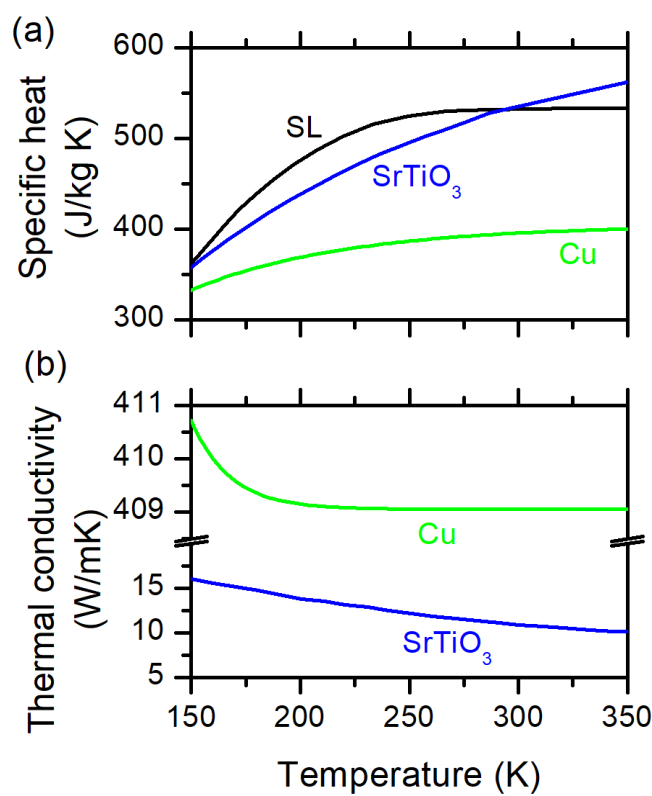

**Figure SM-4:** Temperature dependencies of the material parameters assumed for fitting the thermal model to the experimental data. (a) Shows the specific heat. The generic black curve refers to an SL with m/n=1. Blue and green curves referring to STO and Cu were taken from [70], and [69], respectively. (b) Thermal conductivity of Cu [71] (green) and STO [72] (blue).

E) TEM Characterization

Preliminary TEM characterization was done using a FEI environmental Transmission Electron Microscope (ETEM) Titan 80-300. STEM image simulations were carried using a QSTEM package [73]. Atomic models were built using Vesta software [74]. HR-STEM, iDPC-STEM and EELS experiments were performed at 300 kV using a FEI Titan Themis Z with probe and image aberration correctors, a monochromator and an X-FEG. TEM lamellas were prepared in the [100] and [110] directions using a Thermofischer (former FEI) Helios UC focused ion beam instrument with a beam energy of 30 kV. A final cleaning step was performed at low energy (2 kV). The integrated differential phase contrast (iDPC) STEM technique [75,76] was used to map simultaneously both light and heavy elements at the atomic scale with accuracy comparable with other TEM methods [44].

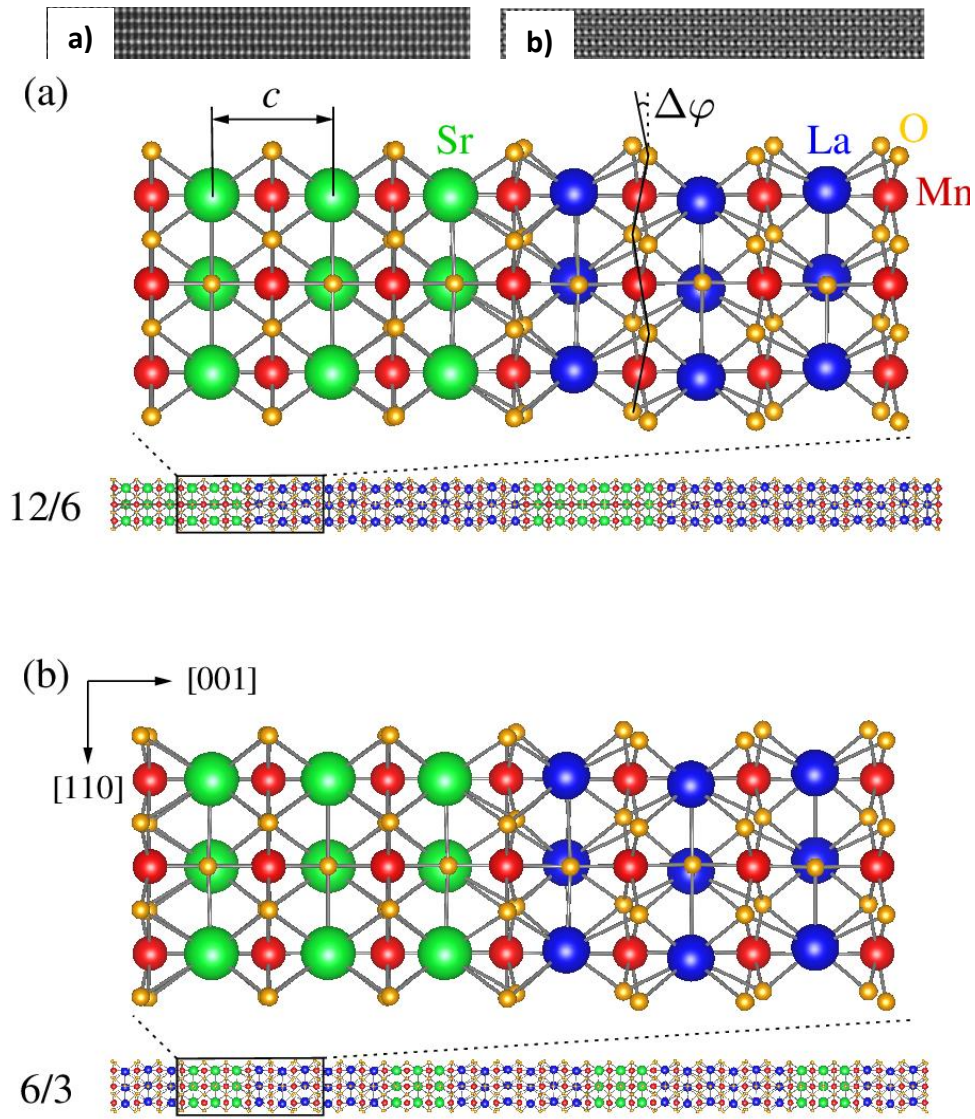

**Figure SM-6:** Atomic structure of t LMO/SMO SLs with m/n=2 optimized from first principles for 12/6 SL (a) and 6/3 SL (b). The tilting of MnO6 octahedra is clearly to seen in both structures.

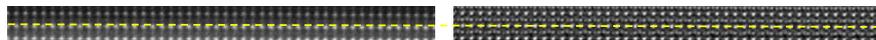

**Figure SM-5**: HAADF-STEM (left) and iDPC-STEM (right) images of 10/5 (a, b), 6/3 (c, d) and 10/10 (e, f) SLs. The contrast between iDPC images of LMO and SMO layers is caused by the deviations of octahedral oxygen rotation angle $\phi_{OOR}$ from 180° in LMO.

**F) Density Functional Theory Calculations**

The structural optimization and all electronic structure calculations were performed from first principles within the density functional theory using the projector augmented-wave (PAW) method as implemented in the VASP code [77]. The exchange-correlation energy was treated using the generalized gradient approximation. The energy cutoff for the plane-wave expansion was set to 450 eV. For the calculated $LMO_m/SMO_n$ SLs, i.e. 6/6, 6/3 and 12/6, we used the $\sqrt{2}\times\sqrt{2}$ in-plane geometry. The structural optimizations were obtained with a conjugate-gradient algorithm and a force tolerance criterion for convergence of 0.015 eV/Å. The Brillouin zone integration was performed using 4×4×1 Monkhorst–Pack k-mesh.

**G) Born-von-Kármán Model (BKM)**

As a starting point of the BKM calculations, we have determined suitable harmonic force constants $K_\alpha$ for the three constituents (cubic SMO, cubic LMO, rhombohedral LMO, see Table SM-III). These constants enter into the models for the SLs. Note that, at the interfaces in the SLs averaged $K_\alpha$ were assumed.

For cubic SMO, we have taken acoustic phonon spectra obtained by DFT calculation [55], which agree well with experimental data [56], to determine the BKM spring constants $K_\alpha$. For cubic LMO, no reliable spectral data was found. Thus, we have considered as a starting point a complex unit cell hosting all 5 different real atomic species as ions ($La^{3+}$, $Mn^{3+}$, $O^{2-}$). We then applied MD potentials [78] to calculate the phonon spectrum in harmonic approximation. For this purpose, the long-range Coulomb interactions was evaluated by means of Ewald summation, short-range van-der-Waals interactions by direct summation. Since such a detailed model is too computationally expensive to construct large SLs, we mapped the results again to the simpler BKM. Our BKM for cubic LMO yields reasonable elastic properties and longitudinal and transverse speeds of sound. [52, 54]. Note that, the maximum LA phonon frequency of the BKM is a bit larger than reported. This is related to the absence of optical modes, which would push the LA modes at the X point further down, and to the compression of the lattice in the out-of-plane direction, compared to a bulk cubic LMO. In case of rhombohedral LMO, we convert elastic constants from Rini et al. [50] into spring constants for the BKM. The resulting elastic and spectral properties comply reasonably with reported properties [52, 54].

| | $K_1$ (N/m) | $K_2$ (N/m) | $K_3$ (N/m) | $v_{LA}$ (m/s) | $v_{TA}$ (m/s) |
|---|---|---|---|---|---|
| SMO (Pm$\bar{3}$m) | 13.13 | 18.94 | 17.51 | 5723 | 4432 |
| LMO (Pm$\bar{3}$m) | 60.08 | 50.24 | 8.16 | 7754 | 4737 |
| LMO (R$\bar{3}$c) | 36.31 | 28.13 | 0 | 5698 | 3217 |

**Table SM-III:** Force constants $K_\alpha$ and acoustic speed of sound along [001] of the BKM for the three different materials in our SLs.

To approximately reproduce the thermal conductivity of SMO at room temperature (293 K) of 4.5 Wm$^{-1}$K$^{-1}$ (see [40] and Figure 2(c)), one can assume for the scattering time parameters $A = 0.96 \cdot 10^{-7}$K²s, and $B = 9.55 \cdot 10^{-4}$K⁴s. The average phonon mean free path at room temperature is $\xi_0 = 24c$ with this parameter set. With $A(T) = A'/T \, e^{\theta_D/3T}$ [58], and $\theta_D = 342$ K (estimated from $v_{LA}$ and $v_{TA}$), also the reported thermal conductivity at 80 K of 11 Wm$^{-1}$K$^{-1}$ (see [40] and Figure 2(c)) can be reproduced by the BKM.

For rhombohedral LMO we have estimated experimentally that, the thermal conductivity shows a small increase with increasing temperature from 100 K to 350 K (see Figure 2(c)). Such a behaviour is anomalous in this temperature range, but known for manganites [61]. It was argued that, the reason is a strong change in the Grüneisen parameter $\gamma$, which enters into the Umklapp scattering time, because $A \sim \gamma^2$ [79]. As a consequence of the anomaly, the conductivity data of LMO includes no clear signature (a maximum at low temperatures) which allows to estimate both Umklapp, and Rayleigh scattering contributions. Therefore, we simply assume a similar amount of Rayleigh scattering as in SMO, and reproduce the room temperature thermal conductivity of 1.3(1) Wm$^{-1}$K$^{-1}$ with $A = 0.35 \cdot 10^{-7}$ K³s. The average phonon mean free path is then $\xi_0 = 18c$.

For cubic LMO, we have estimated experimentally that the thermal conductivity at room temperature should approximately equal that of cubic SMO. Due to a lack of enough low temperature data, we again cannot distinguish between Umklapp and Rayleigh scattering. We again simply assume a similar amount of Rayleigh scattering as in SMO. Therefore, we change only the Umklapp scattering parameter to $A = 3.63 \cdot 10^{-7}$K³s. The average phonon mean free path is then $\xi_0 = 20c$. Comparing cubic and rhombohedral LMO, we find a significant decrease in the Umklapp scattering parameters. This implies that, the lattice anharmonicity strongly increases when the oxygen octrahedra are rotated.

The scattering parameters assumed for fitting the measured thermal conductivities of the SLs are summarized in Table SM-IV. While in principle also for the SLs clearly distinguishing the different contributions is not possible, we have focussed on first-of-all adjusting $B$, and only if necessary $A$. Note that, to crosscheck how strong the determination of the mfp according to Eq. (4) depends on the particular choices for the scattering time parameters $A$ and $B$, we have investigated extremal scenarios, a) by assuming $A \to \infty$, and then finding a suitable $B$ to reproduce $\kappa$, b) by assuming $B \to \infty$, and then finding a suitable $A$ to reproduce $\kappa$, and c) by applying Equation (8) to fit $\kappa$ directly using the mfp as fitting parameter (see Table SM-IV). In any case, we find no significant deviation in the resulting mfp. The reason is that, the integrals in Equation (3) and (A4) are strongly dominated by phonons with $\mathbf{k}||[001]$ whose frequency is around $\omega_0$, as estimated by Equation (A5) (see Table SM-IV). Given the reduced volume thermal conductivity and reduced interfacial conductance, it is surprising that the characteristic mfp in SLs with m/n=2 appears to be larger than in SLs with m/n=1 according to the BKM. This implies that the structural effect on the SL phonon spectrum dominates in the reduction of the volume thermal conductivity $\kappa_v$, overcompensating the increase of $\xi_0$.

|  | $\Lambda/c$ | $\kappa$ (Wm$^{-1}$K$^{-1}$) | $\xi/c$ Eq. (A4) | $A$ ($10^{-7}$ K²s) | $B$ ($10^{-4}$ K⁴s) | $\xi_0/c$ Eq.(4) | $f_0$ (THz) Eq. (A5) |
|---|---|---|---|---|---|---|---|
| SMO (Pm$\bar{3}$m) | 1 | 4.5 [40] | 25 | 0.96 | 9.55 | 24 | 1.73 |
| LMO (Pm$\bar{3}$m) | 1 | 5.1(4) | 19.5 | 3.63 | 9.55 | 20 | 1.43 |
| LMO (R$\bar{3}$c) | 1 | 1.3(1) | 17 | 0.35 | 9.55 | 18 | 1.38 |
| 20/10 SL | 30 | 1.0(1) | 227 | 342.46 | 31 | 250 | 0.24 |
| 12/6 SL | 18 | 0.7(1) | 122 | 9.62 | 1.5 | 122 | 0.16 |
| 10/5 SL | 15 | 0.6(1) | 100 | 9.62 | 0.3 | 101 | 0.14 |
| 8/4 SL | 12 | 0.5(1) | 57.5 | 9.62 | 0.18 | 57 | 0.21 |
| 10/10 SL | 20 | 1.7(3) | 180 | 27.50 | 10.92 | 185 | 0.28 |
| 6/6 SL | 12 | 2.3(1) | 37 | 1.36 | 0.64 | 36 | 0.86 |
| 5/5 SL | 10 | 2.0(3) | 31 | 1.36 | 1.29 | 30 | 1.41 |
| 4/4 SL | 8 | 1.7(4) | 25 | 1.36 | 1.31 | 24 | 1.53 |
| 3/3 SL | 6 | 1.4(1) | 19 | 1.36 | 0.9 | 18 | 1.44 |
| 8/4/4/4 SSL | 20 | 0.0004 | - | 1.36 | 1.29 | 8.3 | 1.59 |
| 10/5/5/5 SSL | 25 | 0.0087 | - | 1.36 | 1.29 | 11.4 | 1.59 |

**Table SM-IV:** Overview about parameters derived from TTR measurements and the BKM. All properties refer to a base temperature of 293 K.